\documentclass[aps,prl,twocolumn,amsmath]{revtex4}

\usepackage{graphicx}
\usepackage{subfigure}
\usepackage{epsfig}
\usepackage{dcolumn}
\usepackage{bm}
\usepackage{color}
\usepackage{xcolor}

\usepackage{amsmath, amssymb, dsfont}

\def \bea{\begin{eqnarray}}
\def \eea{\end{eqnarray}}

\def \ba{\begin{array}{cccc}}
\def \ea{\end{array}}

\begin{document}

\title{Symmetry-protected ideal Weyl semimetal in HgTe-class materials}

\author{Jiawei Ruan$^{1,\ast}$, Shao-Kai Jian$^{2,\ast}$, Hong Yao$^{2,3,\dag}$, Haijun Zhang$^{1,\dag}$, Shou-Cheng Zhang$^{4}$, Dingyu Xing$^1$}

\affiliation{
$^1$National Laboratory of Solid State Microstructures, School of Physics, and Collaborative Innovation Center of Advanced Microstructures, Nanjing University, Nanjing 210093, China\\
$^2$Institute for Advanced Study, Tsinghua University, Beijing 100084, China\\
$^3$Collaborative Innovation Center of Quantum Matter, Beijing 100084, China \\
$^4$Department of Physics, Stanford University, Stanford, California 94305, USA
}

\begin{abstract}
Ideal Weyl semimetals with all Weyl nodes exactly at the Fermi level and no coexisting trivial Fermi surfaces in the bulk, similar to graphene, could feature deep physics such as exotic transport phenomena induced by the chiral anomaly. Here, we show that HgTe and half-Heusler compounds, under a broad range of in-plane compressive strain, could be materials in nature realizing ideal Weyl semimetals with four pairs of Weyl nodes and topological surface Fermi arcs. Generically, we find that the HgTe-class materials with nontrivial band inversion and noncentrosymmetry provide a promising arena to realize ideal Weyl semimetals. Such ideal Weyl semimetals could further provide a unique platform to study emergent phenomena such as the interplay between ideal Weyl fermions and superconductivity in the half-Heusler compound LaPtBi.
\end{abstract}

\maketitle

\begin{figure*}
	\includegraphics[angle=0,width=16cm]{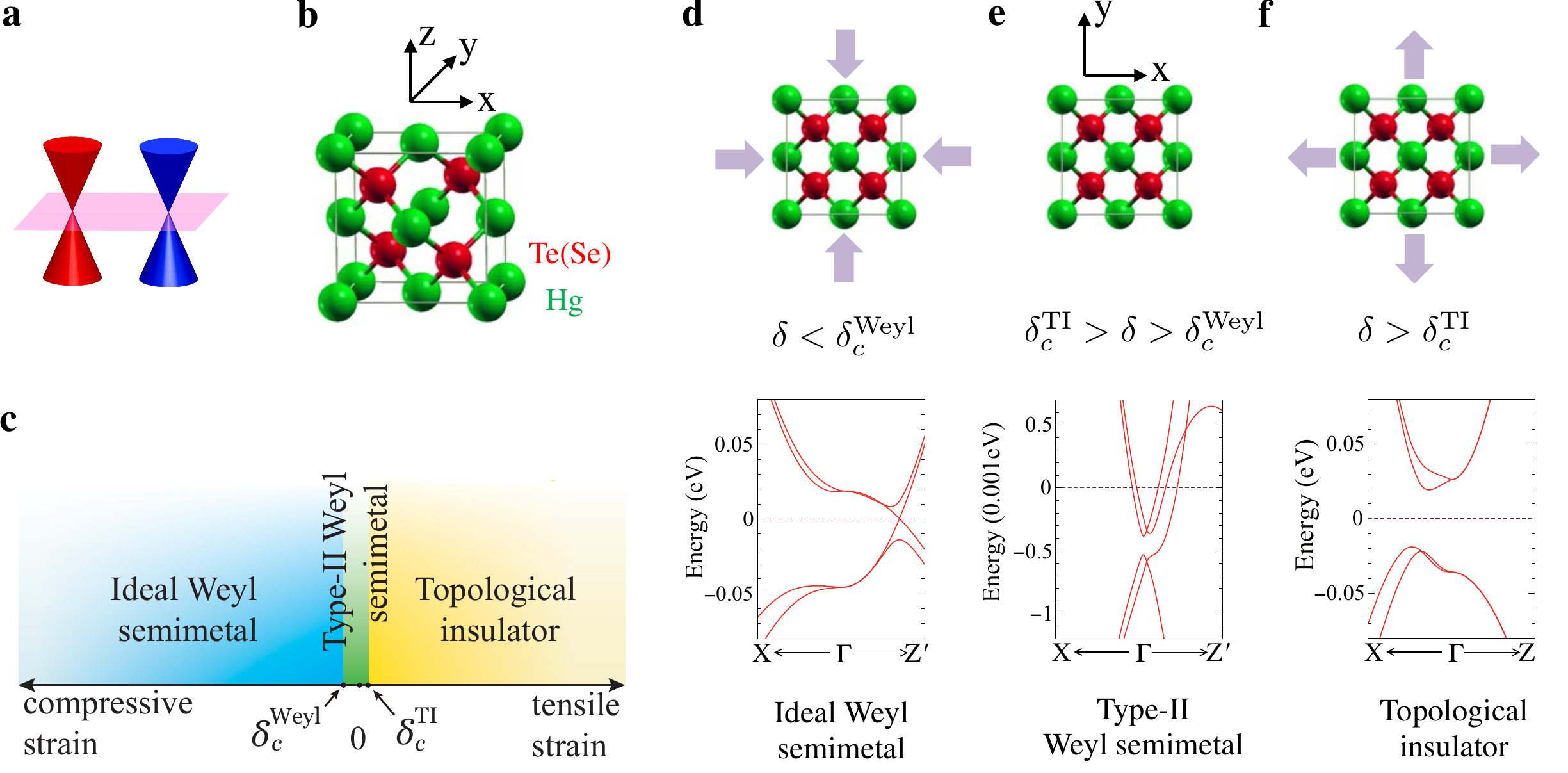}
	\caption{ {$\mid$ \bf Phase diagram of strained HgTe.} ({\bf a}) The Weyl nodes in ideal Weyl semimetals are exactly at the Fermi level without coexisting trivial Fermi pockets. ({\bf b}) The schematic representation of HgTe lattice with a typical zinc-blende structure. ({\bf c}) For a certain range of strain in the $xy$ plane, HgTe has three phases: ideal Weyl semimetals, type-II Weyl semimetals, and topological insulators. ({\bf d}) When the compressive strain is large enough ($\delta < \delta ^{\text{Weyl}}_{c}$), HgTe turns into the ideal Weyl semimetal phase. At stoichiometry, all Weyl nodes are exactly at the Fermi level. ({\bf e}) When the strain is small enough ($\delta ^{\text{Weyl}}_{c} < \delta < \delta ^{\text{TI}}_{c}$), HgTe is in the type-II Weyl semimetal phase. ({\bf f}) When the tensile strain is large enough ($\delta > \delta ^{\text{TI}}_{c}$), HgTe turns into the topological insulator phase. }
\end{figure*}

Weyl fermions were originally introduced as elementary particles in high energy physics more than 80 years ago\cite{weyl1929}. While evidences of Weyl fermions as elementary particles remain elusive, realizing them as low-energy quasiparticles in solids has recently attracted increasing interest \cite{Wan2011, balents2011,xu2011a,burkov2011,yang2011,halasz2012, zhang2014a,liu2014,weng2015,huang2015,hirayama2015,soluyanov2015,lv2015,xu2015,yang2015,xu2015a,lv2015a}, especially after the discovery of topological insulators\cite{Hasan2010,qi2011}.  In solids, Weyl nodes arise as discrete band-crossing points in crystal momentum space near the Fermi level, and Weyl fermions, as quasiparticles with linear dispersions across the Weyl nodes, can be described by the massless Dirac equation\cite{nielsen1983}. Weyl nodes carry a left- or right-handed chirality and they always appear in pairs of opposite chiralities according to the no-go theorem\cite{nielsen1983}. Intriguingly, it has been shown that the chiral anomaly in Weyl semimetals induces many  manifestations in transport properties\cite{nielsen1983, zyuzin2012, wei2012, son2013, liu2013, ashby2013, hosur2013, zhang2015a, ong2015, xindai2015} such as negative magnetoresistance, anomalous Hall effect, and chiral magnetic effects. Another important hallmark of Weyl semimetals is its topologically protected unusual surface states - surface Fermi arcs\cite{Wan2011}.

To realize Weyl semimetals in solids, either the time reversal or lattice inversion symmetry has to be broken\cite{Wan2011,balents2011,xu2011a,burkov2011,yang2011, halasz2012, zhang2014a,liu2014,weng2015,huang2015,hirayama2015,soluyanov2015}. Recently, Weyl nodes were experimentally observed in noncentrosymmetric TaAs-class materials \cite{lv2015,xu2015,yang2015,xu2015a,lv2015a}. It was reported \cite{lv2015,xu2015,yang2015} that these compounds host 24 Weyl nodes which are close to but not exactly at the Fermi level due to the fact that not all Weyl nodes are symmetry-related. Moreover, there are also trivial Fermi pockets at the Fermi level in addition to the Weyl nodes. It is thus desirable to search for ideal Weyl semimetals with all Weyl nodes exactly residing at the Fermi level at stoichiometry composition (as shown in Fig.~1{\bf a}) considering that ideal Weyl semimetals could maximize the potential for transport phenomena induced by the chiral anomaly and that supersymmetry may emerge at the pair-density-wave quantum criticality in an ideal Weyl semimetal\cite{Jian2015}.

In this work, we show that the HgTe-class materials with both band inversion and lattice noncentrosymmetry, including mercury chalcogenide HgX (X=Te, Se)\cite{bernevig2006d} and the multifunctional half-Heusler compounds\cite{Chadov2010,Lin2010,xiao2010},
can realize symmetry-protected ideal Weyl semimetals with four pairs of Weyl nodes under a broad range of in-plane biaxial compressive strain. Note that strain engineering has been successfully employed in condensed matter systems to realize many  physics\cite{ming2015,maier2012}, which makes the experimental realization of strained HgX and half-Heusler compounds highly feasible. For simplicity, we focus on HgTe to discuss concrete results and to illustrate the general guiding principle to realize ideal Weyl semimetals in such class of materials.
\\

\noindent {\bf Results}

\noindent {\bf The effective model of HgTe.} HgTe has a typical zinc-blende structure with the space group F$_{\overline{4}3m}$ (Fig.~1{\bf b}) and it is known to have an inverted band structure \cite{bernevig2006d} such that it is a semimetal and its $\Gamma_8$ bands are half-filled. The $\Gamma_8$ bands at the $\Gamma$ point are fourfold degenerate as $J=3/2$ multiplet. Another important feature of HgTe is its bulk inversion asymmetry (BIA), which plays an essential role in realizing ideal Weyl fermions in strained HgTe as we shall show below. The $\Gamma_8$ bands around the $\Gamma$ point can be effectively described by the following Luttinger Hamiltonian\cite{luttinger1956} plus perturbations due to the BIA:
\begin{eqnarray}
\mathcal{H}_{\text{Luttinger}}({\mathbf k})
    = \alpha_0 {\mathbf k}^2+\alpha_1 ({\mathbf k} \cdot {\mathbf J})^2  + \alpha_2 \sum_{i=1}^3 k_i^2 J_i^2,
\end{eqnarray}	
where $J_i$ are spin-3/2 matrices and $\alpha_i$ are constants characterizing the band structure. The main perturbations induced by BIA are given by $\mathcal{H}_{\text{BIA}}= \alpha[k_x \{J_x,J_y^2-J_z^2\}+ \text{c.p.}]+ \beta[k_x(k_y^2-k_z^2) J_x+ \text{c.p.}]$,
where $\{ \}$ represents anti-commutator, $\text{c.p.}$ means cyclic permutations, and $\alpha, \beta$ are constants characterizing the strength of BIA\cite{winklerbook,dai08prb}. By fitting the first-principle band structures around the $\Gamma$ point, parameters above can be determined: for HgTe, $\alpha_0 \approx 109.8 \textrm{\AA}^2$eV, $\alpha_1 \approx -45.87 \textrm{\AA}^2$eV, $\alpha_2 \approx -19.73 \textrm{\AA}^2$eV, and $\alpha\approx 0.208 \textrm{\AA}$eV.

Because of the BIA, dispersions around the $\Gamma$ point are not purely quadratic and the Fermi level at stoichiometry is slightly above the fourfold degeneracy energy of the $\Gamma_8$ bands\cite{zaheer2013}. Consequently, there exist tiny electron and hole pockets at the Fermi level.  The existence of small Fermi pockets is further indicated by the line crossings between two intermediate bands protected by the crystalline symmetries of the cubic HgTe (Supplementary Figure 1 and Supplementary Note 1). Breaking the crystalline symmetry from T$_d$ to D$_{2d}$ by an in-plane strain can remove the line crossings and render the realization of ideal Weyl semimetals generically inevitable as we show below.

Strain in the $xy$ plane generates a perturbation $\mathcal{H}_{\text{strain}}=- g(J_z^2-\frac54)$, where $g$ depends on the strength of strain and the lattice constant in the $xy$ plane changes to $a=(1+\delta)a_0$. As explained in Supplementary Note 1, $g>0$ ($g<0$) for tensile strain $\delta>0$ (compressive strain $\delta<0$).
The in-plane strain reduces the original cubic symmetry T$_d$ to the tetragonal symmetry D$_{2d}$ in which only two mirror planes (the $k_x=\pm k_y$ planes) survive and the other four are broken. As a result, line-crossings originally protected by the broken mirror symmetries split except at discrete points in the $k_x=0$ or $k_y=0$ plane. The crossing points in the $k_x=0$ or $k_y=0$ plane are robust because these planes respect a special symmetry ${\mathrm C}_{2T}= {\mathrm C}_2 \cdot T$, where C$_2$ is the two-fold rotation along the $x$ or $y$-axis and $T$ is the time-reversal transformation (see Supplementary Note 1 for details). In total, there are eight Weyl nodes with four in the $k_x=0$ or $k_y=0$ plane, respectively.

\noindent {\bf The phase diagram of strained HgTe. } Under a sufficiently small strain, these Weyl nodes are type-II\cite{soluyanov2015} (see also Supplementary Figure 2); they are close to but not exactly at the Fermi level and there are also other coexisting trivial Fermi pockets. Note that tensile and compressive strains have qualitatively different effect in moving these type-II Weyl nodes. Under a large enough tensile strain, the eight Weyl points annihilate with each other in the $k_x\!=\!\pm k_y$ planes, leading to a strong topological insulator (TI), as shown in Fig.~1{\bf f}, which agrees with previous theoretical calculations\cite{dai08prb} and experimental observations\cite{brune11prl}. Intriguingly, under an increasing compressive strain, these type-II Weyl nodes quickly evolve to type-I Weyl nodes, shown in Fig.~1{\bf d}, all of which lie exactly at the Fermi level leading to an ideal Weyl semimetal at stoichiometry, similar to graphene! Based on the {\bf k}$\cdot${\bf p} Hamiltonian $\mathcal{H}_{k\cdot p} = \mathcal{H}_{\text{Luttinger}} + \mathcal{H}_{\text{BIA}} + \mathcal{H}_{\text{strain}}$, we obtain the whole phase diagram, schematically shown in Fig.~1{\bf c}, where the strained HgTe is a strong TI when $\delta>\delta_c^\textrm{TI}$, and is an ideal Weyl semimetal when  $\delta<\delta_c^\textrm{Weyl}$.  In the ideal Weyl semimetal phase, HgTe has eight Weyl nodes at $(\pm k_x^\ast,0,\pm k_z^\ast)$ and $(0,\pm k_y^\ast,\pm k_z^\ast)$, as schematically shown in Fig.~2.

\noindent {\bf The analysis of topological properties.} Now, we use the {\bf k}$\cdot${\bf p} theory to analyze topological properties of ideal Weyl nodes. We first consider the effect of the strain and treat the BIA as a perturbation. It is straightforward to show that $\mathcal{H}_0 \equiv \mathcal{H}_{\text{Luttinger}} +\mathcal{H}_{\text{strain}}$ hosts two Dirac points in the $k_z$ axis. The BIA perturbation splits each Dirac point into four Weyl nodes. The effective Hamiltonian around one of the Weyl points ($k_x^\ast,0,k_z^\ast$) is described by the Weyl equation:
\bea
\mathcal{H}_{\text{Weyl}}= \sum_{i=x,y,z} v_i k_i \sigma^i,
\eea
where $v_i$ are Fermi velocities given in Supplementary Note 3. For HgTe, we find that the Weyl node located at ($k_x^\ast, 0, k_z^\ast$) is right-handed, from which the chirality of other Weyl nodes can be derived since all of them are related by symmetry. The locations and chiralities of the eight ideal Weyl nodes are shown in Fig.~2.

\begin{figure}[t]
	\includegraphics[width=6.0cm]{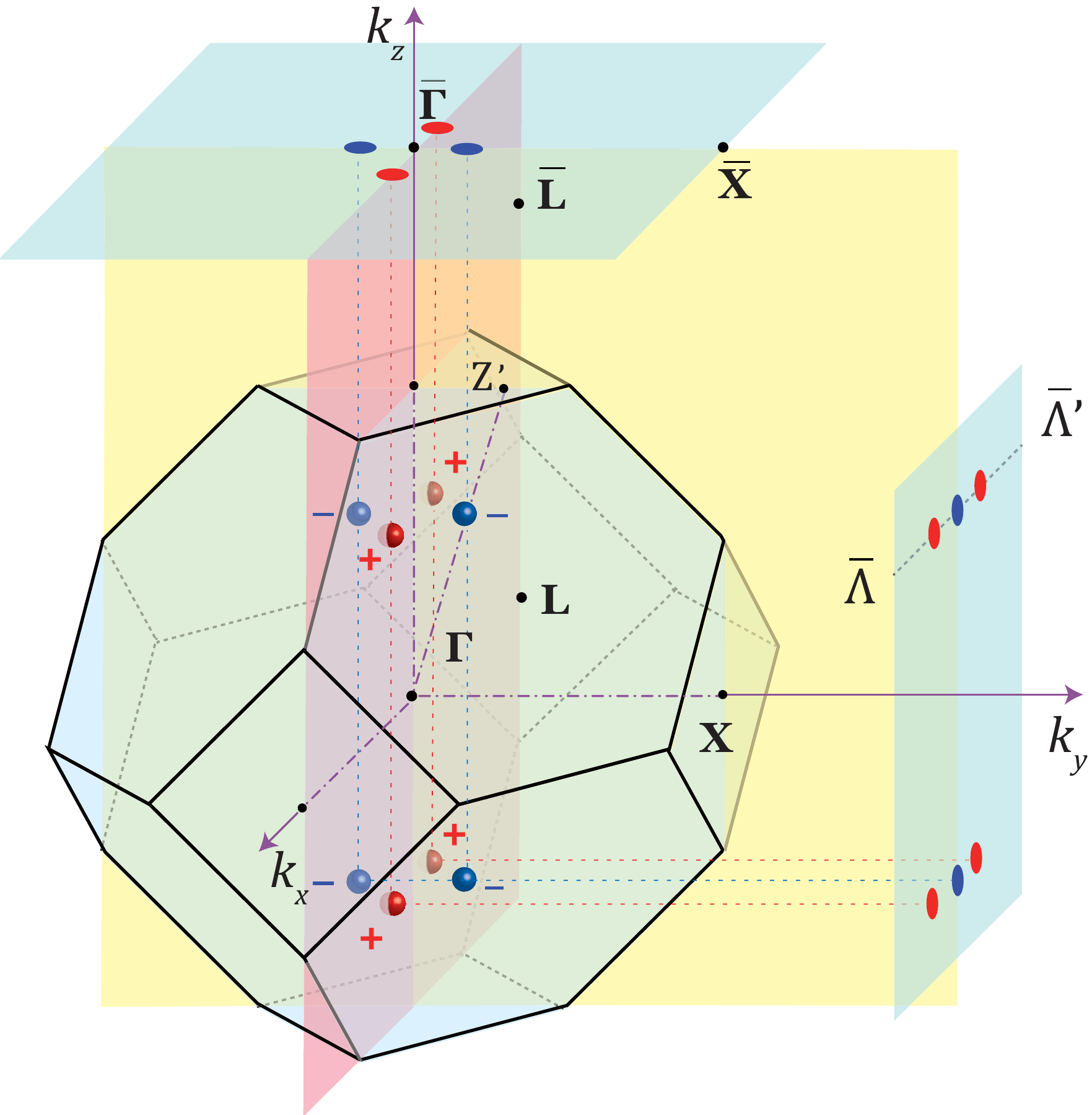}
	\caption{{$\mid$ \bf Schematic of Weyl points in Brillouin zone.} The bulk Brillouin zone (BZ), and (001) and (010) surface BZs of HgTe. In the ideal Weyl semimetal phase of the strained HgTe, there are four pairs of ideal Weyl nodes in the bulk BZ, schematically shown as red (chirality $+1$)  and blue (chirality $-1$) circles. The pink (yellow) plane is for the $k_y=0$  ($k_x=0$) plane. The projections of bulk Weyl nodes onto the (001) and (010) surface BZs are shown; there are four gapless points in the (001) surface BZ but six gapless points in the (010) surface BZ. }
\end{figure}

One hallmark of Weyl semimetal is the existing of topologically protected surface Fermi arcs. In the ideal Weyl semimetal phase, the electronic states in the $k_z=0$, $k_x=k_y$, and $k_x=-k_y$ planes are all gapped so that the $Z_2$ topological-invariant in any of these two-dimensional time-reversal-invariant subsystems is well-defined; these $Z_2$-invariants are all nontrivial because of the band inversion at the $\Gamma$ point. Consequently, these subsystems carry gapless helical edge modes, which have important implications to possible Fermi arc patterns. Combined with the known chirality of different Weyl nodes, it is expected that the Fermi arcs form a closed circle in (001) surface Brillouin zone (BZ) and that there are open Fermi arcs on the (100) or (010) surface BZ.

\begin{figure}[t]
	\includegraphics[angle=0,width=8.5cm]{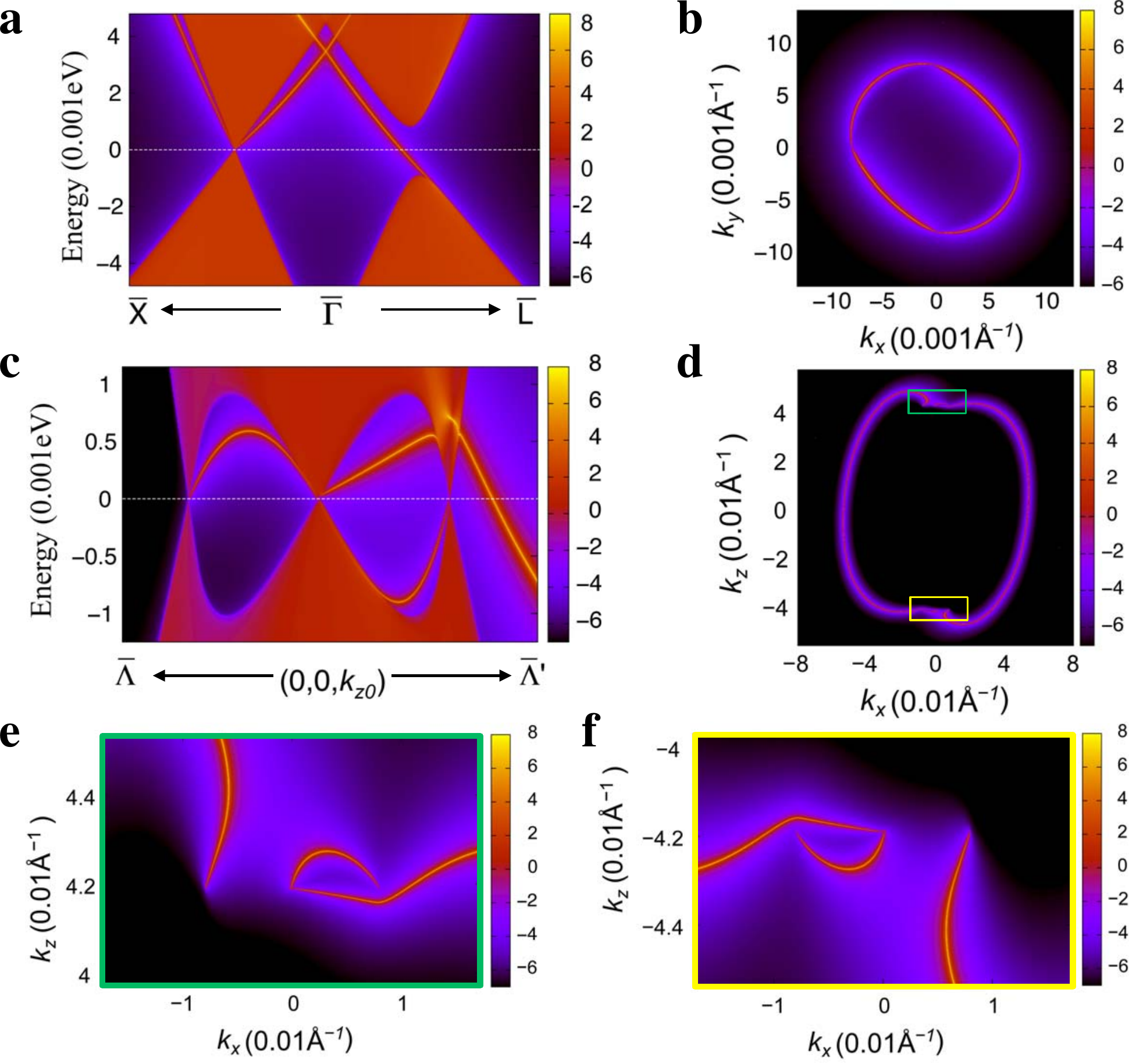}
	\caption{{$\mid$ \bf Surface states and Fermi arcs.} Electronic structure of HgTe with surfaces in the ideal Weyl semimetal phase. ({\bf a}) Band structure projected onto the (001) surface BZ. The red lines denote the surface states. ({\bf b}) Surface Fermi arcs in the (001) surface BZ form a closed loop, because each gapless point has the monopole charge of $+2$ or $-2$. ({\bf c}) Band structure projected onto the $(010)$ surface BZ. ({\bf d}) Surface Fermi arcs in the $(010)$ surface BZ. ({\bf e,f}) To see Fermi arcs clearly, the Fermi surfaces around the Weyl nodes are zoomed in.}
\end{figure}

\noindent {\bf The first-principles calculations.} To confirm the results predicted in the effective ${\mathbf k}$$\cdot$${\mathbf p}$ theory above, we carry out first-principles calculations on HgTe. For the compressive in-plane strain with $a=0.99a_0$ and $c=1.02a_0$, while the fourfold degeneracy of the $\Gamma_8$ band at the $\Gamma$ point is lifted to open a gap, the two intermediate bands touch at eight discrete points: ($\pm k_x^\ast,0,\pm k_z^\ast$) and ($0,\pm k_y^\ast, \pm k_z^\ast$) with $k_x^\ast=k_y^\ast\approx 0.0073$\AA$^{-1}$ and $k_z^\ast\approx 0.042$\AA$^{-1}$, schematically shown in Fig.~ 2. These eight gapless points are exactly the Weyl nodes predicted by the effective $k$$\cdot$$p$ model above. Importantly, these Weyl nodes all exactly locate at the Fermi level without coexisting with any trivial bands. It precisely realizes an ideal Weyl semimetal! On the contrary, with the tensile strain with $a=1.01a_0$ and $c=0.98a_0$, a full band gap opens in the entire BZ, which results in a strong TI because of the band inversion between the $\Gamma_6$ and $\Gamma_8$ bands. Qualitative difference between tensile and compressive in-plane strains in HgTe is due to the different response of $p_{x,y}$ and $p_z$ orbitals to the strain.

\begin{figure*}[t]
	\includegraphics[angle=0,width=13.0cm]{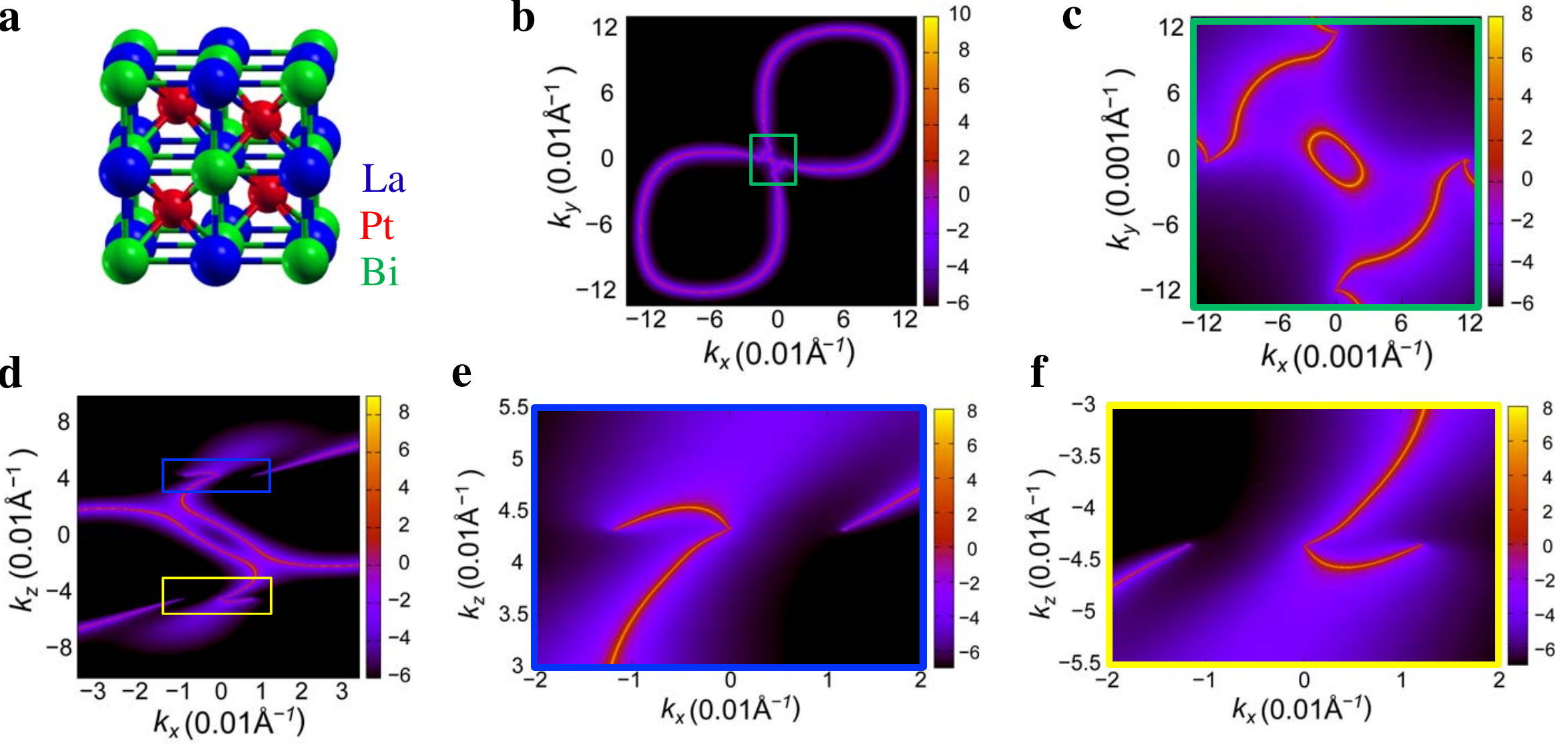}
	\caption{{$\mid$ \bf Surface states and Fermi arcs of strained LaPtBi. }Electronic structure of LaPtBi with surfaces in the
ideal Weyl semimetal phase. ({\bf a}) The stuffed zinc-blende structure of LaPtBi. ({\bf b}) Surface Fermi arcs in the (001) surface BZ also forms a closed loop, though the pattern is different from that of HgTe. ({\bf c}) The surface states around $\overline{\Gamma}$ are zoomed in. ({\bf d,e,f}) Fermi surfaces in the (010) surface BZ exhibit Fermi arcs in ({\bf d}), and the marked regions are zoomed in in ({\bf e}) and ({\bf f}), respectively. }
\end{figure*}

A key feature of the Weyl semimetal is its unusual surface Fermi arcs, which start from one Weyl node and end at another with opposite chirality\cite{Wan2011}. Fig.~3{\bf a} shows the projected bands along $\overline{X}$-$\overline{\Gamma}$-$\overline{L}$ in the (001) surface BZ and surface states can be seen clearly. As the gapless point ($0, \overline{k_y^\ast}$) in the $\overline{X}$-$\overline{ \Gamma}$ line, projected from the two Weyl nodes ($0, k_y^\ast,\pm k_z^\ast$), has monopole charge $-2$, two Fermi arcs have to connect to this point. Similar consideration can be applied to other gapless points ($0, -\overline{k_y^\ast}$) and ($\pm \overline{k_x^\ast},0$) in the (001) surface BZ. The Fermi arcs form a closed loop, as shown in Fig.~3{\bf b}, which is consistent with the nontrivial $Z_2$ topological-invariant defined in the two-dimensional planes of $k_x=\pm k_y$. Along the bulk-gapped line of $\overline{ \Gamma}$-$\overline{ L}$, there is a single edge mode across the Fermi level, consistent with the topological properties of Fermi arcs forming a loop on the (001) surface BZ.

The Fermi arcs on the (010) surface is even more interesting. From the projected bands shown in Fig.~3{\bf c}, we can see that a single chiral edge mode comes out of the gapless points $(\overline{k_x^\ast},\overline{k_z^\ast})$ and two chiral edge modes come out of the gapless point ($0,\overline{k_z^\ast}$), which are consistent with their monopole charges $+1$ and $-2$, respectively. The corresponding Fermi surface is shown in Fig.~3{\bf d}, and the marked regions are zoomed in in Fig.~3{\bf e,f}, where discontinuous Fermi arcs are clearly seen. The Fermi arc originating from the point ($-\overline{k_x^\ast},k_z^\ast$) and terminating at the point ($0,-\overline{k_z^\ast}$) spans across a significant portion of the BZ (about $9\%$ of reciprocal lattice constant), which is of great advantage for experimental detection such as angle resolved photoemission spectroscopy (ARPES). The Fermi arc pattern may change upon varying chemical environment on the surface, but the parity of number of Fermi arcs is topologically stable.

The analysis of realizing ideal Weyl fermions in strained HgTe leads to a general guiding principle to search for ideal Weyl semimetals in noncentrosymmetric materials with nontrivial band inversion. Noncentrosymmetric half-Heusler compounds with band inversion are another family of such materials which could also realize ideal Weyl semimetals under a compressive in-plane strain. For illustration, we perform first-principles calculations on LaPtBi (Fig.~4{\bf a}) with a compressive in-plane strain of $a=0.99a_0$ and $c=1.02a_0$. Indeed, it also shows eight ideal Weyl nodes at ($\pm k_x^\ast,0,\pm k_z^\ast$) and ($0,\pm k_y^\ast, \pm k_z^\ast$) with $k_x^\ast=k_y^\ast=0.013$\AA$^{-1}$ and $k_z^\ast=0.043$\AA$^{-1}$.
Similar to HgTe, the surface Fermi arcs also forms a closed loop in the (001) surface BZ, as shown in Fig.~4{\bf b,c}.
The Fermi arcs in the (010) surface are shown in Fig.~4{\bf d}, whose marked regions are zoomed in in Fig.~4{\bf e,f}.
\\

\noindent{\bf Discussion}

\noindent We have shown that applying in-plane strain in HgTe-class materials can give rise to ideal Weyl semimetals with only four pairs of Weyl nodes exactly locating at the Fermi level. Experimentally, such strain in the $xy$ plane may be effectively obtained by growing samples on substrates with a smaller lattice constant. For example, materials with zinc-blende structure such as  GaSb, InAs, CdSe, ZnTe, and HgSe are promising substrates for growing HgTe to achieve the desired in-plane compressive strain. The locations of Weyl nodes for various strength of strains in HgTe and LaPtBi are calculated and summarized in Supplementary Table 1 (see also Supplementary Note 2). The predicted Weyl nodes and surface Fermi arcs can be directly verified by ARPES measurements. Moreover, transport experiments can be used to detect unusual bulk magneto-transport properties induced by the chiral anomaly of ideal Weyl fermions\cite{hosur2013}.

Ideal Weyl semimetals predicted in the strained HgTe-class materials are interesting on their own. It further provides a perfect platform to study the interplay between ideal Weyl fermions and other exotic phenomena, especially in half-Heusler compounds, including superconductivity in  LaPtBi\cite{goll2008}, magnetism in GdPtBi, and heavy fermion behaviour in  YbPtBi\cite{canfield1991}. Interestingly, spacetime supersymmetry may also emerge at the superconducting quantum critical point in ideal Weyl fermions\cite{Jian2015}. This  class of ideal Weyl semimetals thus opens a broad avenue for fundamental research of emergent physics as well as potential applications of low-power quantum devices.
\\

\noindent{\bf Methods}

{\footnotesize \noindent{\bf The first-principles calculations.} The first-principle calculations are carried out in the framework of the Perdew-Burke-Ernzerhof-type generalized gradient approximation of the density functional theory through employing the BSTATE package\cite{fang2002} with the plane-wave pseudo-potential method. The kinetic energy cutoff is fixed to 340eV, and the {\bf k}-point mesh is taken as 16$\times$16$\times$16 for the bulk calculations. The spin-orbit coupling is self-consistently included. The experimental lattice constants are used with $a_0=6.46$\AA~for HgTe and 6.829\AA~for LaPtBi. In order to simulate the uniaxial strain along the [001] direction, we fix the experimental volume but change the ration $a/c$, where $a$ is the lattice constant in the $x$-$y$ plane and $c$ is the lattice constant along $z$ axis (the [001] direction). Without loss of generality, we take the parameters $a=0.99a_0$ and $c=1.02a_0$ for the tensile strain and  $a=1.01a_0$ and $c=0.98a_0$ for the compressive strain. To exhibit unique surface states and Fermi arcs on the surface, we employ maximally localized Wannier functions\cite{marzari1997,souza2001} to first obtain the {\it ab initio} tight-binding model of the bulk HgTe and LaPtBi and then pick these bulk hopping parameters to construct the tight-binding model of the semi-infinite system with the (001) or (010) surface as the boundary. The surface Green's function of the semi-infinite system, whose imaginary part is the local density of states to obtain the dispersion of the surface states, can be calculated through an iterative method\cite{Zhang2009a}.}

\noindent {\bf Acknowledgement}\\
\noindent We thank Yulin Chen and Xiangang Wan for helpful discussions and appreciate G. Yao for technical supports and J. Sun for sharing his computing resource. H.Y. is supported in part by the National Thousand-Young-Talents Program and by the NSFC under Grant No. 11474175 at Tsinghua University. H.J.Z is supported by the Scientific Research Foundation of Nanjing University (020422631014) and the National Thousand-Young-Talents Program. S.C.Z. is supported by  the Department of Energy, Office of Basic Energy Sciences, Division of Materials Sciences and Engineering, under contract DE-AC02-76SF00515 and by FAME, one of six centers of STARnet, a Semiconductor Research Corporation program sponsored by MARCO and DARPA.\\

\noindent {\bf Author contributions}\\
\noindent {\footnotesize H.Y. and H.Z. designed this project. S.-K.J. and H.Y. performed the analysis of the effective {\bf k$\cdot$p} model. J.R. and H.Z performed the first-principles calculations. S.-C.Z and D.X. supervised the project.  H.Y. and H.Z. wrote the manuscript with the input from all authors.} \\

\noindent {\bf Additional information}\\
{\footnotesize \noindent Supplementary information is available in the online version of the paper. {\bf Reprints and permissions} information is available online at www.nature.com/reprints.\\

\noindent J.W.R. and S.-K.J. contributed equally to this work. Correspondence and requests for materials should be addressed to H.Y. (yaohong@tsinghua.edu.cn) or H.J.Z. (zhanghj@nju.edu.cn).\\

\noindent {\bf Competing financial interests:} The authors declare no competing financial interests.}

\renewcommand{\figurename}{{\bf Supplementary Figure}}
\renewcommand{\tablename}{{\bf Supplementary Table}}
\renewcommand{\thetable}{\arabic{table}} 

\begin{widetext}

\begin{figure}[t]
	\centering
	\includegraphics[angle=-90,width=12cm]{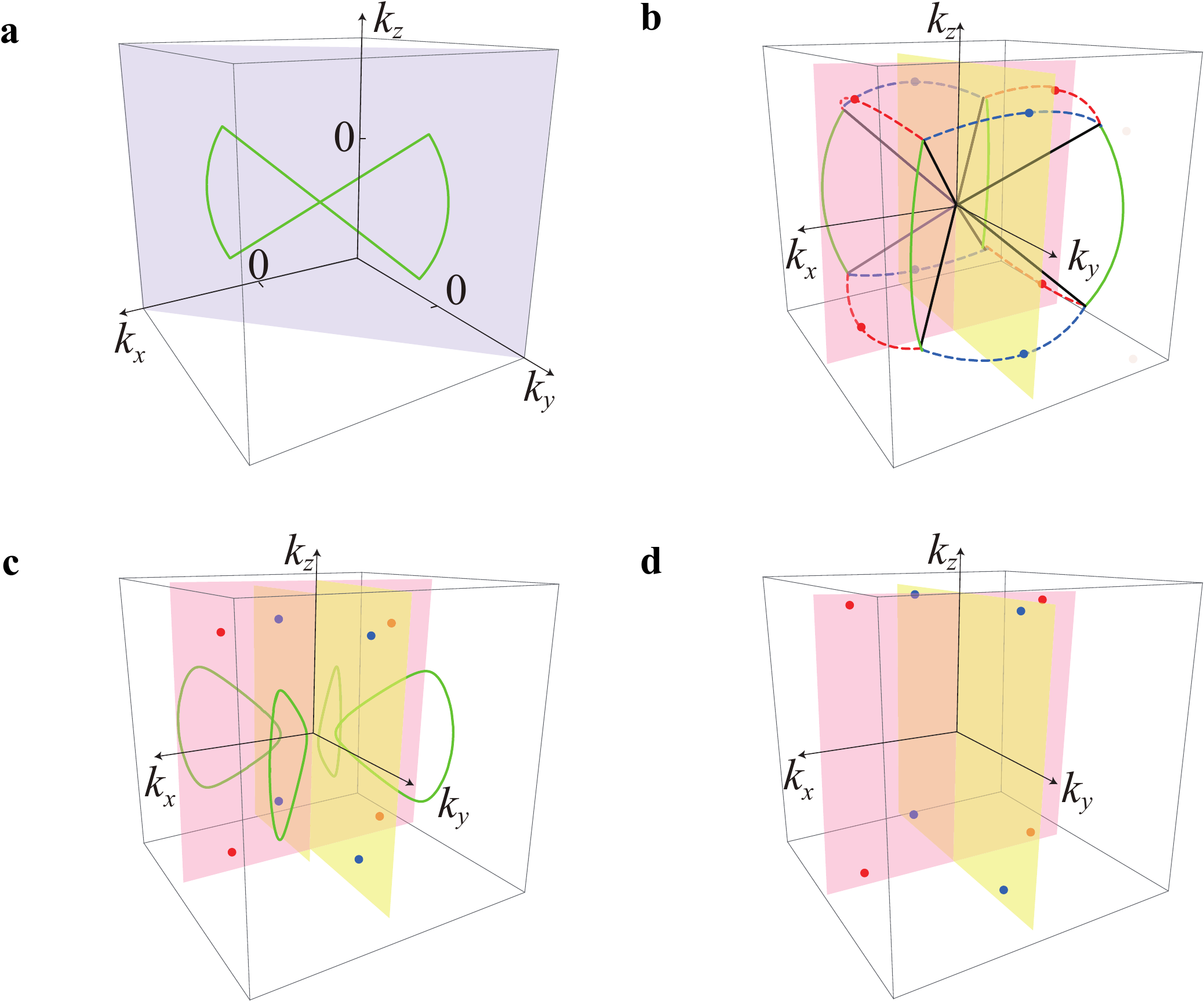}
	\caption{{$\mid$ \bf The characterization of band crossings in HgTe with and without applied strain. } ({\bf a}) The green line represents the line nodes in the $k_x=-k_y$ mirror plane of HgTe without strain. ({\bf b}) The line node structure without strain in the full bulk BZ. The blue, red, and green lines represent line nodes lying in the mirror plane but not along any diagonal direction. The black solid lines label line nodes in diagonal directions. The red (blue) dashed lines cross the $k_y=0$ ($k_x=0$) plane where the crossing points are indicated by red (blue) points. These dashed line nodes, except those crossing points, split once strain is applied due to the breaking of their mirror symmetries. ({\bf c}) After applying a small strain, the line nodes in the $k_x=\pm k_y$ mirror planes gradually shrink, as shown by the green lines. The red and blue points represent the Weyl nodes in the $k_x=0$ and $k_y=0$ planes, protected by C$_{2T}$ symmetry. ({\bf d}) When the compressive strain exceeds a critical value, i.e. $\delta<\delta^{\mathrm Weyl}_c$, there are no line nodes any more and the Weyl nodes, indicated by the red and blue points, are type-I such that they are all located exactly at the Fermi level and the system is in the ideal Weyl semimetal phase. } \label{line_node}
\end{figure}


\begin{figure}
    \centering
\includegraphics[angle=-90,width=12cm]{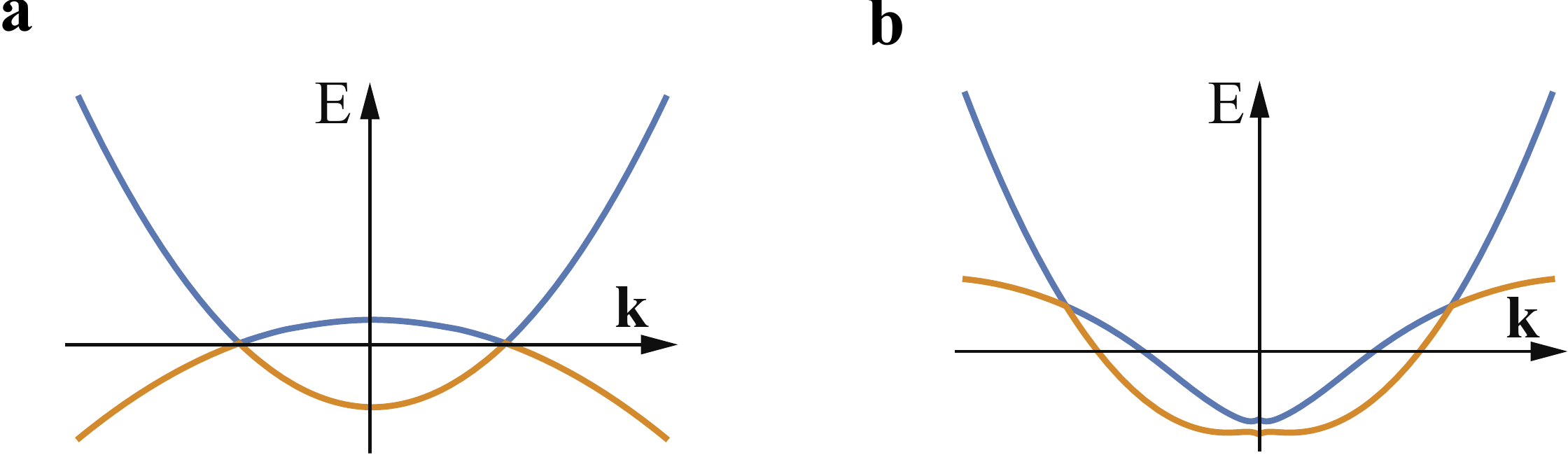}
    \caption{{$\mid$ \bf Schematic dispersions of the two types of Weyl nodes.} ({\bf a}) ideal Weyl semimetals and ({\bf b}) type-II Weyl semimetals. }\label{dispersion}
\end{figure}


\begin{table}
\caption{The location of Weyl nodes $(k^\ast_x,0,k^\ast_z)$ under different in-plane strains for (a) HgTe and (b) LaPtBi. Here $a$ and $c$ denote the lattice constants after applying the in-plane strain, and $a_0$ denotes the experimental lattice constant of HgTe or LaPtBi.}\label{strain}
\subtable[~HgTe]{\begin{tabular}{|c| c| c |c|}
\hline
~~~~$a/a_0$~~~~ & ~~~~$c/a_0$~~~~ & $~~k^\ast_x ({\mathrm \AA}^{-1})~~$ &~~ $k^\ast_z({\mathrm \AA}^{-1})$~~ \\
\hline
0.995  & 1.01 & .0069 & .0294 \\
\hline
0.990 & 1.02 & .0073 & .0417 \\
\hline
0.985 & 1.03 & .0077 & .0513 \\
\hline
0.980 & 1.04 & .0080 & .0593 \\
\hline
0.976 & 1.05 & .0085 & .0671 \\
\hline
0.971 & 1.06 & .0086 & .0737 \\
\hline
\end{tabular}}~~~~~~~~~~~~~~
\subtable[~LaPtBi]{
\begin{tabular}{|c| c| c |c|}
\hline
~~~~$a/a_0$~~~~ & ~~~~$c/a_0$~~~~ & $~~k^\ast_x ({\mathrm \AA}^{-1})~~$ &~~ $k^\ast_z({\mathrm \AA}^{-1})$~~ \\
\hline
0.995  & 1.01 & .0098 & .0306 \\
\hline
0.990 & 1.02 & .0133 & .0434 \\
\hline
0.985 & 1.03 & .0166 & .0541 \\
\hline
0.980 & 1.04 & .0198 & .0639 \\
\hline
0.976 & 1.05 & .0231 & .0729 \\
\hline
0.971 & 1.06 & .0267 & .0817 \\
\hline
\end{tabular}}
\end{table}


\section{Supplementary Information}

\noindent{\bf Supplementary Note 1: Systematic evolution of the band structure under strain}\\
\noindent As mentioned in the main text, the HgTe-class materials which carry T$_d$ point group symmetry, do not respect inversion symmetry. The inversion asymmetry in these systems results in qualitatively different band structures compared to those with inversion symmetry, e.g., $\alpha$-Sn. Instead of a simple quadratic band touching point (at $\Gamma$) exactly at Fermi level, two intermediate bands of the four $\Gamma_8$ bands in HgTe-class materials touch along a line node in the six mirror planes. These line nodes are protected by mirror symmetry even though they are generically away from the Fermi level. As a consequence, the HgTe-class materials have small electron and hole pockets, even at stoichiometry.

To illustrate these line nodes, we diagonalize Hamiltonian in the mirror plane, e.g., the $k_x=-k_y$ plane as shown by the shaded plane in Supplementary Figure 1{\bf a}. The unstrained Hamiltonian is given in the main text, i.e., $\mathcal{H}_{\text{unstrained}} =\mathcal{H}_{\text{Luttinger}} +\mathcal{H}_{\text{BIA}}$. For simplicity,  only the linear term in $\mathcal{H}_{\text{BIA}}$ is taken into consideration and higher-order terms in $\mathcal{H}_{\text{BIA}}$ won't affect the results we obtain in this section qualitatively. The line nodes lying in the $k_x=-k_y$ mirror plane is given by the following equation:
\bea
	 2(c_1^2-c_2^2) k_z^2-3|\alpha^2-2 c_1^2 k_z^2|   	+(2c_2^2-8c_1^2) k_y^2+3 \alpha^2 = 0.
\eea
where $c_i$ is defined in Supplementary Note 3. This bow-shape line node is shown in Supplementary Figure \ref{line_node}{\bf a} by a green line. There are six symmetry-related mirror planes and each of them contains such a bow-shape line node. Two line nodes overlap each other along the diagonal direction where their mirror planes intersect, as indicated by the black lines in Supplementary Figure \ref{line_node}{\bf b}. Since the system has a C$_3$ symmetry along the diagonal direction, the two intermediate bands form a two-dimensional irreducible representation that explains the overlaps of two line nodes along this direction. The red , blue or green lines in Supplementary Figure \ref{line_node}{\bf b} indicate the parts of the line nodes not along the diagonal directions.

Upon applying strain in the $xy$ plane, the crystalline symmetry of the material is lowered from T$_d$ to D$_{2d}$. The mirror symmetries in the $k_x=\pm k_z$ and $k_y=\pm k_z$ planes are broken, while those in $k_x=\pm k_y$ planes survive. As a result, for small strain the line nodes located at $k_x=\pm k_y$ plane survive as they are protected by the unbroken mirror symmetry, even though they contract as the applied strain increases. The evolution of the line nodes is shown in Supplementary Figure \ref{line_node}{\bf c} where the closed green lines indicate the shrinking line nodes under a small strain. When the applied strain exceeds critical value, these line nodes eventually disappear.

On the other hand, the line nodes in $k_x=\pm k_z$ and $k_y=\pm k_z$ planes are split immediately , upon applying strain except the eight discrete points. These eight discrete points are the intersecting points between the $k_y=0$ ($k_x=0$) plane and the line nodes in the $k_x=\pm k_z$ ($k_y=\pm k_z$) planes, as shown by red or blue color points in Supplementary Figure \ref{line_node}{\bf b}. As mentioned in the main text, these eight discrete points are protected from gapping out by C$_{2T}\equiv$ C$_2 \cdot T$ symmetries. Applying strain only gradually shift these points in the $k_y=0$ or $k_x=0$ planes. Supplementary Figure \ref{line_node}{\bf c} shows these discrete points upon applying a small strain. These eight discrete points are actually Weyl points, which reside in the $k_x=0$ or $k_y=0$ planes protected by the C$_{2T}$ symmetry (see Supplementary Note 3 for details).

When the strain is sufficiently small but finite, these points are type-II Weyl points. These type-II Weyl points in the $k_y=0$ plane shift to $|k_z|>|k_x|$ region if compressive strain is applied while they shift to $|k_z|<|k_x|$ region if tensile strain is applied, and Weyl points in the $k_x=0$ have the similar behavior related by symmetry. Thanks to this qualitatively different response between tensile and compressive strain, the system evolve into different phases at sufficiently large strain. When the tensile strain increases, these type-II Weyl points first move to the $k_x$ or $k_y$ axis in the $k_z=0$ plane, then move within the $k_z=0$ plane (since there is C$_{2T}$ symmetry in the $k_z=0$ plane), and finally they annihilate one another with opposite chiralities in $k_z=0, k_x= \pm k_y$ lines when the tensile strain reaches a critical value. When the strain exceeds the critical value, the system enters into a strong topological insulator phase with nontrivial $Z_2$ topological-invariant.

On the other hands, when the compressive strain increases from zero, the eight type-II Weyl nodes shift towards a larger $|k_z|$ direction. When the compressive strain exceeds a critical value, all trivial Fermi surface vanishes and the eight type-II Weyl nodes evolve to type-I ideal Weyl nodes located exactly at the Fermi level, as shown in Supplementary Figure \ref{line_node}{\bf d}.
\\

\noindent {\bf Supplementary Note 2: The Weyl nodes under different strains}\\
\noindent In this section, we calculate locations of ideal Weyl nodes for different in-plane compressive strains for both HgTe and the half-Heusler compound LaPtBi, as shown in Supplementary Table 1. The Weyl nodes move slowly towards larger momentum points in both $k_x$ and $k_z$ directions for increasing strain. As the Weyl nodes can only be pair-annihilated in the $k_z=\pi$ plane, the slow motion of Weyl nodes with increasing strain indicates that the ideal Weyl semimetal phase is stable under a broad range of strain. Indeed, for the large strain of $a/a_0=0.971$ and $c/a_0=1.06$, the Weyl nodes are still far away from the $k_z=\pi$ plane. Moreover, the larger separation between Weyl nodes in momentum space under increasing in-plane compressive strain can make the observation of them by Angle resolved photoemission spectroscopy (ARPES) experiments easier.
\\

\noindent {\bf Supplementary Note 3: The effective k$\cdot$p theory at finite strain}\\
\noindent In this section, we use following Gamma matrices \cite{murakami2004}:
\bea
	\Gamma^1= \frac{1}{\sqrt{3}} \{J_y, J_z\},~
	\Gamma^2=  \frac{1}{\sqrt{3}} \{J_z, J_x\},~
	\Gamma^3= \frac{1}{\sqrt{3}} \{J_x, J_y\},
	\Gamma^4 = \frac{1}{\sqrt{3}} (J_x^2- J_y^2),~
	\Gamma^5= J_z^2- \frac{5}{4}
\eea
to write the Luttinger Hamiltonian:
\bea
\mathcal{H}_{\text{Luttinger}}({\bf k})
    = c_0 {\bf  k}^2+c_1 \sum_{i=1}^3 d_i \Gamma^i  + c_2 \sum_{i=4}^5 d_i \Gamma^i,
\eea
where $d_1({\bf  k})\!=\!\sqrt{3}k_y k_z$, $d_2({\bf k})\!=\!\sqrt{3} k_x k_z$, $d_3({\bf k})\!=\!\sqrt{3} k_x k_y$, $d_4({\bf  k})\!=\! \frac{\sqrt{3}}{2}(k_x^2-k_y^2)$, $d_5({\bf  k})\!=\! \frac{1}{2}(2k_z^2-k_x^2-k_y^2)$ and $c_0\!=\!\alpha_0+\frac54(\alpha_1+\alpha_2)$, $c_1\!=\!\alpha_1$, $c_2\!=\!(\alpha_1+\alpha_2)$.

Now, we explore the behavior of HgTe under sufficiently large strain (exceeds the critical value $g^{\text{Weyl}}_c$) using the effective {\bf k}$\cdot${\bf p} theory. From Clebsch-Gordan coefficients, we know
\bea
|\frac32,\frac32\rangle &=& |p_x+ip_y,\uparrow\rangle,~|\frac32,-\frac32\rangle = |p_x-ip_y,\downarrow\rangle,  \\
|\frac32,\frac12\rangle &=& \sqrt{\frac13}|p_x+ip_y,\downarrow\rangle+\sqrt{\frac23}|p_z,\uparrow\rangle,~
|\frac32,-\frac12\rangle = \sqrt{\frac13}|p_x-ip_y,\uparrow\rangle+\sqrt{\frac23}|p_z,\downarrow\rangle,
\eea
where $|\frac32,J_z\rangle$ labels the wave function of a single electron with $J_z=\pm \frac32,\pm \frac12$. Electrons with $J_z=\pm 3/2$ have definite orbital angular momentum $l_z=\pm 1$ while electrons with $J_z=\pm 1/2$ are superpositions of the wave functions with $l_z=\pm 1$ and $l_z=0$. Since compressive strain in the $xy$ plane shortens the lattice distance along the $x$ and $y$ directions, the wave functions with orbital angular momentum $l_z=\pm 1$ get more overlapped than that of $l_z=0$, and their energy shift is larger than that of $l_z=0$. This indicates that the main effect induced by the strain in the $xy$ plane is given by the following perturbation:
\bea
	\mathcal{H}_{\text{strain}}&=&- g (J_z^2-\frac54)= -g\Gamma^5,
\eea
where $g$ is related to the strength of applied strain $\delta$. Moreover, we obtain $g<0$ for compressive strain ($\delta<0$) and $g>0$ for tensile strain ($\delta>0$).

We shall focus on the case that $g<g^{\text{Weyl}}_c$ here. For this case, as explained in the main text, treating $\mathcal{H}_0 \equiv \mathcal{H}_{\text{Luttinger}}+\mathcal{H}_{\text{strain}}$ as unperturbed Hamiltonian and the BIA part $\mathcal{H}_\textrm{BIA}$ as perturbation is a better way to characterize the band features around Weyl nodes. The quadratic touching point in Luttinger Hamiltonian $\mathcal{H}_\textrm{Luttinger}$ is split into two Dirac points locating at $k_z$ axis for compressive strain because $\mathcal{H}_\textrm{strain}$ has only D$_{4h}$ symmetry which is lower than the O$_h$ symmetry of the Luttinger Hamiltonian. Specifically, the dispersion is given by $E=c_0 k^2 \pm\sqrt{c_1^2(d_1^2+d_2^2+d_3^2)+c_2^2 d_4^2+ (c_2 d_5+ g)^2)}$, leading to two Dirac points  at $(0,0, \pm \sqrt{g/c_2})$. Expanding the Hamiltonian around the touching point $(0,0,\sqrt{g/c_2})$, we obtain the Dirac Hamiltonian:
\bea
\mathcal{H}_{\text{Dirac}}({\bf k}) &\equiv& v_z' k_z+ v_\perp(k_y \Gamma^1+ k_x \Gamma^2) + v_z k_z \Gamma^5,  \label{Dirac}
\eea
where $v_z'=2c_0\sqrt{\frac{g}{c_2}}, v_\perp= -\sqrt{\frac{3g c_1^2}{c_2}}, v_z=-2\sqrt{g c_2}$. These Dirac points are protected by inversion symmetry, time reversal symmetry, as well as the S$_4(z)$ symmetry of $\mathcal{H}_0$. In HgTe, inversion symmetry is actually broken which has important consequences even though the breaking is weak. As a result of BIA, linear, cubic, as well as higher-order terms compatible with the T$_d$ symmetry are allowed in the Hamiltonian and we treat them as perturbations.

We first consider the effect of the linear term in $\mathcal{H}_\textrm{BIA}$\cite{winklerbook}:
\bea
	\mathcal{H}_{\text{linear}} &=& \alpha \Big[ (\frac{-\sqrt{3}}{2} \Gamma^{15}+ \frac{3}{2} \Gamma^{14}) k_x+ (\frac{-\sqrt{3}}{2} \Gamma^{25}- \frac{3}{2} \Gamma^{24}) k_y + \sqrt{3} \Gamma^{35} k_z \Big].
\eea
For simplicity, we approximate $\mathcal{H}_\textrm{linear}$ by its form at the Dirac point $(0,0,\sqrt{g/c_2})$ and denote it as $\mathcal{H}_{\text{m}}=m \Gamma^{35}$, where $m=\alpha \sqrt{3 g/c_2}$. In the presence of $\mathcal{H}_\textrm{m}$, the low-energy dispersion is given by $E_k=v_z'k_z \pm \sqrt{(v_\perp \sqrt{k_x^2+k_y^2} \pm m)^2 + v_z^2 k_z^2}$. It is easy to see that the Dirac points are split and lead to two doubly degenerate line nodes satisfying the two equations: $k_x^2+k_y^2= (m/v_\perp)^2$ and $k_z=\pm\sqrt{g/c_2}$. In general, line nodes in 3D momentum space are not stable against further generic perturbations. For instance, the cubic term in $\mathcal{H}_\textrm{BIA}$ given by $\mathcal{H}_\textrm{cubic}=\beta(\{k_x,k_y^2-k_z^2\} J_x+ \text{cp})$, where $\beta$ is a constant that describes the strength of cubic term, can split the line nodes. Interestingly, eight discrete gapless points, which are the crossing points of original line nodes with the $k_x=0$ or $k_y=0$ plane, survive from generic perturbations in $\mathcal{H}_\textrm{BIA}$. These stable gapless discrete points in the $k_x=0$ or $k_y=0$ plane are protected by a special symmetry C$_{2T}= $C$_2 \cdot T$ that forms little group of these planes which we shall explain below.

We now consider the $k_y=0$ plane to explain how its C$_{2T}$ symmetry can protect a gapless Weyl point in this plane. The $k_y=0$ plane respects the C$_{2T}$ symmetry which is given by
\bea
	{\mathrm C}_{2T} \propto e^{-i\pi J_y} \cdot e^{i\pi J_y} K=K,
\eea
where $K$ is the complex conjugation operator, an anti-unitary transformation. After expanding the Hamiltonian, i.e., $\mathcal{H}_\text{Dirac}+\mathcal{H}_m$, at one of the four crossing points $(\frac{m}{v_\perp},0,\sqrt{\frac{g}{c_2}})$ in the $k_y=0$ plane, we obtain an effective  Hamiltonian:
\bea
\mathcal{H}({\bf  k})=v_z' k_z+ v_\perp (k_x \Gamma^2+k_y \Gamma^1)+ v_z k_z \Gamma^5+ m(\Gamma^2+ \Gamma^{35}).
\eea
Since two intermediate bands of four bands touches, we implement a unitary transformation $U$ that diagonalizes $\Gamma^2$ and $\Gamma^{35}$ simultaneously to project the Hamiltonian to those bands. The resultant Hamiltonian in the $k_y=0$ plane reads ($\mathcal{H}_U=U\mathcal{H} U^\dag$):
\bea
	\mathcal{H}_U(k_y=0) = \left( \ba -v_\perp k_x - 2m & v_z k_z & 0 & 0 \\ v_z k_z & v_\perp k_x + 2m & 0 & 0 \\ 0 & 0 & v_z k_z & - v_\perp k_x \\ 0 & 0 & -v_\perp k_x & -v_z k_z \ea \right),
\eea
which has a simple block-diagonal form. It is easy to see that the low-energy sector is given by second block and it can be written compactly as:
\bea
	\mathcal{H}'_U(k_y=0)= v_z k_z \sigma^z- v_\perp k_x \sigma^x.
\eea
Moreover, the C$_{2T}$ operator does not change after this unitary transformation, i.e. $U{\mathrm C}_{2T}U^\dag= {\mathrm C}_{2T} \propto K$. An immediate consequence is that the gapless point locating at $k_y=0$ plane is protected: any gap-opening term proportional to $\sigma^y$ is not allowed because it breaks the C$_{2T}$ symmetry. Namely, because the Hamiltonian respects this special C$_{2T}$ symmetry, the eight Weyl points in the $k_x=0$ and $k_y=0$ planes are stable against all possible BIA and strain perturbations if they are not too strong. These perturbations can only shift these Weyl points in the $k_x=0$ and $k_y=0$ plane.

We now include the cubic term $\mathcal{H}_\text{cubic} \equiv \beta[k_x(k_y^2-k_z^2) J_x+ \text{c.p.}] $ in $\mathcal{H}_\text{BIA}$ to obtain the effective Hamiltonian of the Weyl points with dispersions away from the C$_{2T}$-planes. Expanding it at the gapless point $(\frac{m}{v_\perp},0,\sqrt{\frac{g}{c_2}})$, we obtain a linear term in $k_y$: $\frac{2}{\sqrt{3}}v_y J_yk_y$, where $J_i$ is the angular momentum operator, and $v_y=\sqrt{3} \beta \left( \frac{g}{c_2}- (\frac{m}{v_\perp})^2 \right)=\sqrt{3} \beta (\frac{g}{c_2}-\frac{\alpha^2}{c_1^2})$. Projecting this linear terms into the low-energy subspace of two intermediate bands around the gapless Weyl point, we obtain the Weyl Hamiltonian:
\bea
	\mathcal{H}_{\text{Weyl}} &=& \sum_i v_i k_i \sigma^i.
\eea
where $v_x\!=\!\sqrt{\frac{3g \alpha_1^2}{\alpha_1+\alpha_2}}$, $v_y\!=\!-\sqrt{3} \beta (\frac{g}{\alpha_1+\alpha_2}- \frac{\alpha^2}{\alpha_1^2})$, and $v_z\!=\!-2\sqrt{g (\alpha_1+\alpha_2)}$ and it is exactly the Weyl Hamiltonian appeared in the main text. For HgTe, we find that the velocity $v_y$  is  negative because $\beta>0$. So this Weyl node is right-handed as mentioned in the main text.
\\


\end{widetext}
\end{document}